\documentclass[conference]{IEEEtran}
\IEEEoverridecommandlockouts
\usepackage{cite}
\usepackage{amsmath,amssymb,amsfonts}
\usepackage{algorithmic}
\usepackage{graphicx}
\usepackage{textcomp}
\usepackage{xcolor}
\usepackage{booktabs}
\usepackage{enumitem}
\usepackage{float}
\usepackage[hidelinks]{hyperref}

\usepackage{titlesec}
\titlespacing*{\section}{0pt}{*1}{*0.4}
\titlespacing*{\subsection}{0pt}{*0.6}{*0.2}
\titlespacing*{\subsubsection}{0pt}{*0.4}{*0.1}

\begin{document}

\title{KATANA: A Fast, Low-Power Mapping of Kalman Filters onto Edge NPUs for Real-Time Tracking\vspace{-15pt}}

\author{
\IEEEauthorblockN{Bodhisatwa Kundu$^{1}$, Anish Rooj$^{1}$, Sumit Saha$^{1}$, Abhradeep Sarkar$^{1}$, Arghadip Das$^{2}$, Arnab Raha$^{3}$, Mrinal K. Naskar$^{1}$}
\IEEEauthorblockA{$^{1}$Jadavpur University, West Bengal, India
$^{2}$Purdue University, IN, USA
$^{3}$Intel Corporation, CA, USA}
\vspace{-60pt}}

\maketitle

\begin{abstract}
State estimation is the closed-loop core of every real-time tracking system, from radar surveillance, missile guidance, and counter-unmanned aerial vehicle (UAV) defense to autonomous driving and robotics. All of these deployments run on edge platforms: defense systems mount on vehicles, drones, and interceptors far from fixed infrastructure, while civilian pipelines live on cars, drones, robots, and handheld devices, where every additional watt of compute erodes mission duration or operational range. Two hard constraints follow: each new measurement must be fused before the next control cycle (a few milliseconds, scaled by the number of tracked targets), and the total compute must fit within a strict battery and thermal power envelope. The Linear and Extended Kalman Filters (LKF, EKF) are the dominant estimators on this class of system, but today they execute almost exclusively on the CPU, which serializes multi-object tracking (MOT) updates, or on custom FPGA/ASIC accelerators that lengthen design cycles and add silicon area. Contemporary AI personal computer (AI-PC) system-on-chip (SoC) silicon, such as the Intel Core Ultra Series 1 and Series 2, already integrates a low-power, data-parallel Neural Processing Unit (NPU) alongside the CPU and GPU; we therefore ask whether the Kalman filter (KF) can be mapped onto this existing matrix engine to meet real-time and low-power budgets simultaneously. Such a mapping avoids a dedicated accelerator and keeps the CPU and GPU free for their primary workloads. We present \textbf{KATANA}, a novel NPU-aware optimization framework that delivers the first end-to-end mapping of the LKF and EKF onto a commercial NPU, together with the first cross-platform (CPU, GPU, NPU) characterization of these filters on shipping AI-PC silicon. KATANA applies three algebraic graph rewrites: subtract-to-add reformulation via a precomputed negative-projection matrix $\mathbf{H}_{\text{neg}}$, static-shape tensor fusion, and block-diagonal batched parallelization, so that 100\% of operations execute on the Data Processing Unit (DPU) matrix engine. On the Series 2, the optimized batched EKF reaches \textbf{223.35 frames per second (FPS)} at \textbf{13.43~W} active power and the LKF reaches \textbf{408.73 FPS} at \textbf{14.05~W}, delivering up to a \textbf{97.9\%} reduction in dynamic energy versus the CPU implementation.
\end{abstract}

\section{Introduction}
Real-time state estimation is the closed-loop heartbeat of modern tracking systems. In defense, ground-based and airborne radars are mounted on vehicles, ships, and aircraft far from fixed infrastructure \cite{b1,b2}, while counter-UAV and missile-guidance loops run on drones and interceptors that themselves move at high speed. In civilian deployments, the same tracking workloads run on cars (for advanced driver assistance), drones, robots, and handheld devices \cite{b3,b4}, where the tracker has to live on-device for latency and connectivity reasons. All of these are edge platforms: battery-powered and either fanless or weight-constrained, so every additional watt of compute erodes mission duration or operational range, while the control loop must still close within milliseconds of each new measurement and scaled by the number of tracked targets. Across all of these settings, the LKF and its non-linear counterpart, the EKF, have been the dominant estimators for over six decades \cite{b5,b6}: optimal in the minimum-mean-square-error (MMSE) sense, recursive (hence streaming-friendly), and structured around dense linear algebra that any modern compute substrate should be able to execute.

The hardware to do so already exists. Contemporary client SoCs, such as the Intel Core Ultra Series 1 and Series 2, Apple M-series, and Qualcomm Snapdragon X, are now heterogeneous, integrating a CPU, a GPU, and a dedicated NPU optimized for low-power AI inference. Yet, despite this hardware diversity, the KF today runs almost exclusively on the CPU in commodity software stacks, or on custom FPGA/ASIC accelerators in dedicated tracking systems \cite{b7}. Three problems follow. \textit{First}, MOT with $N$ independent filters serializes on the CPU, capping throughput and burning power. \textit{Second}, designing a dedicated accelerator extends time-to-deployment and adds die area on top of an already-busy SoC. \textit{Third}, the on-die NPU, a data-parallel matrix engine designed for sustained low-power operation, sits idle whenever no neural workload is feeding it.
\textit{Mapping the KF to this otherwise-idle NPU yields a double dividend: it avoids a separate accelerator and keeps the CPU and GPU free for their primary general-compute and graphics workloads, so the tracker coexists with rather than monopolizes the SoC, making the end-to-end system more responsive.}

\begin{figure}[t]
    \centering
    \includegraphics[width=0.7\columnwidth]{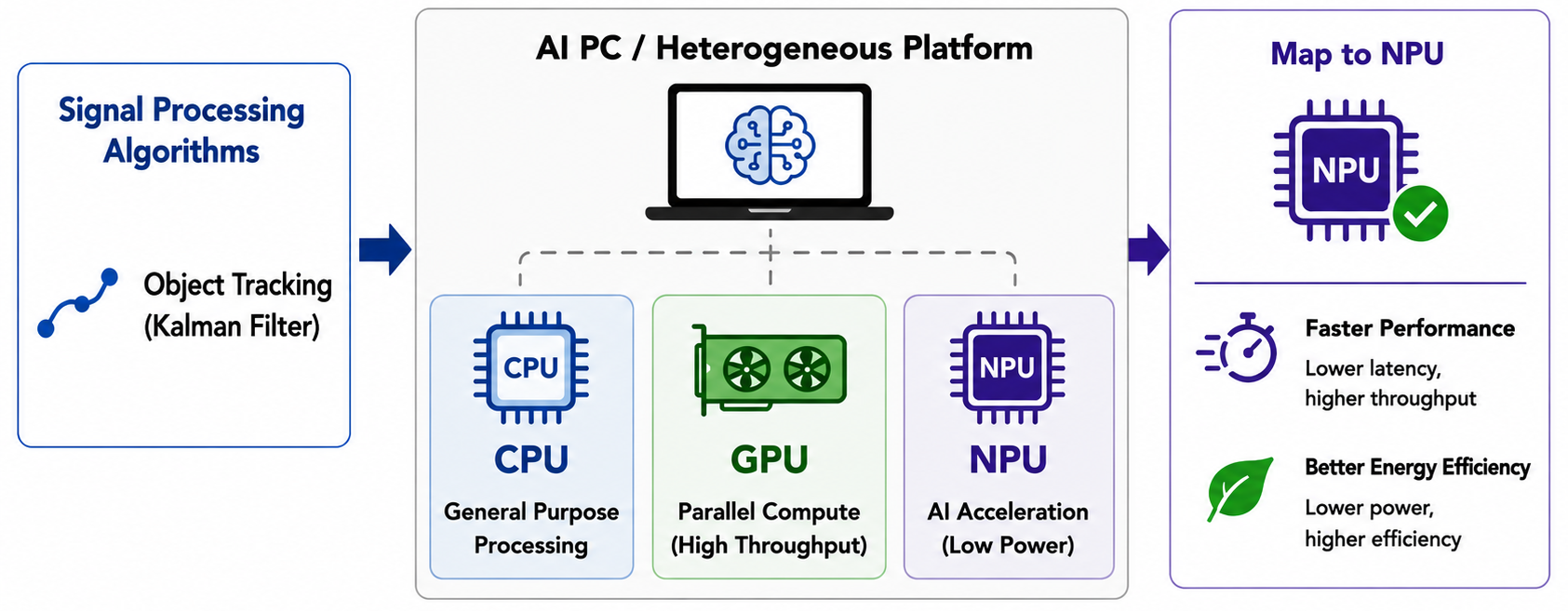}
    \caption{Overview. KATANA targets a classical signal-processing workload, Kalman-filter-based object tracking (left), on heterogeneous AI-PC SoCs (centre) and maps the recursive estimator onto the on-die NPU (right) for faster and more energy-efficient execution.}
    \label{fig:npu_arch_overview}
\end{figure}

This work therefore asks the natural question: \textit{can the KF be mapped onto the existing AI-PC NPU and made real-time at low active power, without custom silicon?} Fig.~\ref{fig:npu_arch_overview} captures the overall premise: take a classical signal-processing algorithm such as KF-based tracking, target a heterogeneous AI-PC platform, and offload the recursive estimator to the on-die NPU. The challenge, however, is that NPUs are designed around dense multiply-accumulate (MAC) dataflow on the DPU; any operation that falls outside that pattern (Subtract, Reshape, Transpose, Gather) is routed to a scalar Digital Signal Processor (DSP) and forces costly DPU$\leftrightarrow$DSP context switches that can consume 10--30\% of inference time on the small kernels typical of KF tracking.

Building on the above motivation, we present KATANA, a novel NPU-aware optimization framework for mapping traditional state estimators onto AI-PC silicon. Our contributions are:
\begin{itemize}
    \item We present the \textbf{first cross-platform (CPU, GPU, NPU) characterization} of the LKF and EKF inference on two consecutive AI-PC generations (Series 1 and Series 2).
    \item We introduce three \textbf{NPU-aware algebraic graph rewrites}: a precomputed negative-projection matrix $\mathbf{H}_{\text{neg}}$ that converts subtract-heavy innovations into DPU-native adds, static-shape tensor fusion that removes runtime Reshape/Transpose nodes, and block-diagonal batched parallelization that packs $N$ independent filters into one inference call. Together, these rewrites move 100\% of operations onto the DPU.
    \item We demonstrate that the optimized NPU pipeline meets real-time latency budgets within a sustained \textbf{13--14~W} active envelope, sustains \textbf{$>$200~FPS} multi-filter throughput on the Series 2, and reduces dynamic energy by up to \textbf{97.9\%} versus CPU execution.
    \item We validate end-to-end on a live video stream, where the NPU-resident LKF and EKF consume \textbf{$<$1\%} of a 33~ms frame budget at 30~FPS, leaving the CPU and GPU free for detection and downstream analytics.
\end{itemize}

\section{Background}
We briefly review the algorithmic primitive we wish to accelerate (Section~II-A) and the NPU substrate on which we will run it (Section~II-B). Together, these two pieces frame the optimization problem addressed in Section~\ref{sec:method}.

\subsection{Kalman Filtering for State Estimation}
The KF is a recursive estimator for the hidden state of a noisy dynamical system \cite{b5,b6}. The LKF assumes $\mathbf{x}_k = \mathbf{F}\mathbf{x}_{k-1} + \mathbf{w}_{k-1}$ and $\mathbf{z}_k = \mathbf{H}\mathbf{x}_k + \mathbf{v}_k$ with Gaussian process and measurement noise $\mathbf{w}_k, \mathbf{v}_k$, and alternates a prediction step (propagating the state and covariance through $\mathbf{F}$) with an update that corrects via the innovation $\mathbf{y}_k = \mathbf{z}_k - \mathbf{H}\hat{\mathbf{x}}_{k|k-1}$ scaled by the Kalman gain $\mathbf{K}_k = \mathbf{P}_{k|k-1}\mathbf{H}^T(\mathbf{H}\mathbf{P}_{k|k-1}\mathbf{H}^T+\mathbf{R})^{-1}$. The EKF keeps the same linear-gain structure but linearizes the dynamics and observation maps about the current estimate via the Jacobians $\mathbf{F}_k, \mathbf{H}_k$. Each recursion therefore reduces to a chain of dense matrix multiplications plus a single inversion.

\subsection{NPU Microarchitecture and Execution Flow}
\begin{figure}[t]
    \centering
    \includegraphics[width=\columnwidth]{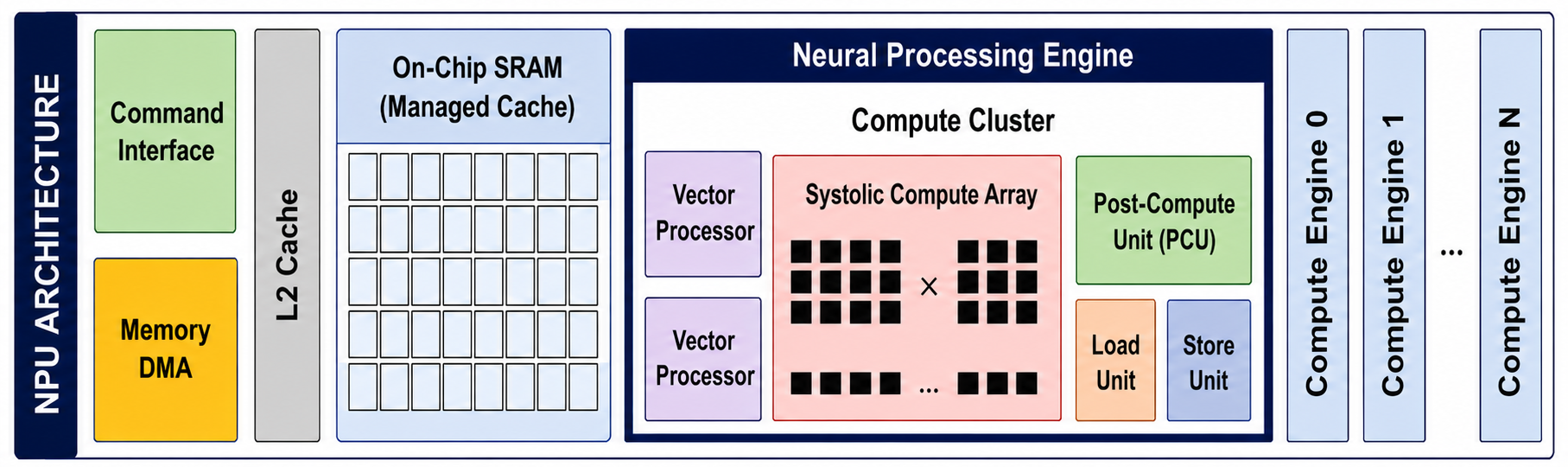}
    \caption{NPU microarchitecture: Command Interface, on-chip SRAM with managed L2 cache, and a Compute Cluster of Compute Engines, each combining a systolic DPU, two DSP units, a PCU, and Load/Store units.}
    \label{fig:npu_arch_detail}
\end{figure}

Fig.~\ref{fig:npu_arch_detail} shows the Intel NPU on the Series 1 and Series 2 SoCs: a Command Interface, a managed on-chip SRAM with DMA, and a Compute Cluster of identical Compute Engines that each pair a Systolic DPU (dense MAC) with two Vector DSP units, a Post-Compute Unit, and Load/Store units \cite{b8,b9}; the same architecture and its five-stage tensor-execution pipeline are described in detail in \cite{b14}. Only one property matters for the rest of the paper: the DPU dominates throughput and energy efficiency on dense GEMM, while any op routed to the DSP serializes with the DPU pipeline and triggers a DPU$\leftrightarrow$DSP context switch whose fixed cost is significant for the small tensors typical of KF tracking. Maximizing DPU occupancy is therefore the central design problem for Section~\ref{sec:method}.

\section{Related Work}
Our work sits at the intersection of two research threads: hardware acceleration of state estimators, and algorithm--NPU co-design for non-CNN workloads.

\textbf{Hardware acceleration of KF tracking.} FPGA implementations of multi-dimensional KFs for object tracking achieve deterministic latency but require lengthy RTL design cycles and carry higher static power than current SoC fabrics \cite{b7}. GPU implementations exploit batch parallelism in many-target settings but exceed the thermal envelopes of fanless and battery-powered edge devices, and lower-precision strategies such as stochastic computing \cite{b10} reduce per-operation energy in principle without addressing the DSP$\leftrightarrow$DPU context-switch cost on heterogeneous AI accelerators.

\textbf{NPU co-design for non-CNN workloads.} Modern NPUs increasingly host workloads beyond convolutional networks: FlexNPU offers a dataflow-flexible substrate for energy-efficient edge inference \cite{b9}, and recent frameworks have mapped LLMs \cite{b11}, GNNs \cite{b12}, SSMs \cite{b13}, and Hyena/Kolmogorov--Arnold Networks \cite{b14} onto resource-constrained NPUs. A consistent finding is that every successful NPU mapping requires algorithm-level restructuring to align the workload with the matrix-engine dataflow.

\textbf{Gap.} Classical signal-processing kernels (KFs in particular) have not, to our knowledge, been mapped to a commercial NPU, and no cross-platform (CPU, GPU, NPU) characterization exists for the LKF and EKF on shipping AI-PC silicon. KATANA fills this gap.

\section{NPU-Aware Graph Optimization for Kalman Filters}
\label{sec:method}
Our design goal follows directly from Section~II-B: every operation in the LKF and EKF prediction-update recursion must execute on the DPU, with zero fall-back to the DSP, so that the NPU's GEMM pipeline is the only critical path. All filters are authored in PyTorch, exported to ONNX, and compiled to each target backend through Intel OpenVINO 2024.5 \cite{b8,b4}. Three graph rewrites, applied prior to compilation, transform a naive ONNX export into a pure-DPU graph. Fig.~\ref{fig:netron_single_column} traces the resulting Netron graphs for the LKF (top panel) and EKF (bottom panel) through four stages: \textit{Baseline}, \textit{Optimization-1} (subtract elimination), \textit{Optimization-2} (static-shape tensor fusion), and \textit{Batched} (block-diagonal parallelization).

\subsection{Optimization Pipeline Overview}
\begin{figure}[t]
\centering
\begin{minipage}{\columnwidth}
    \centering
    \includegraphics[width=\columnwidth]{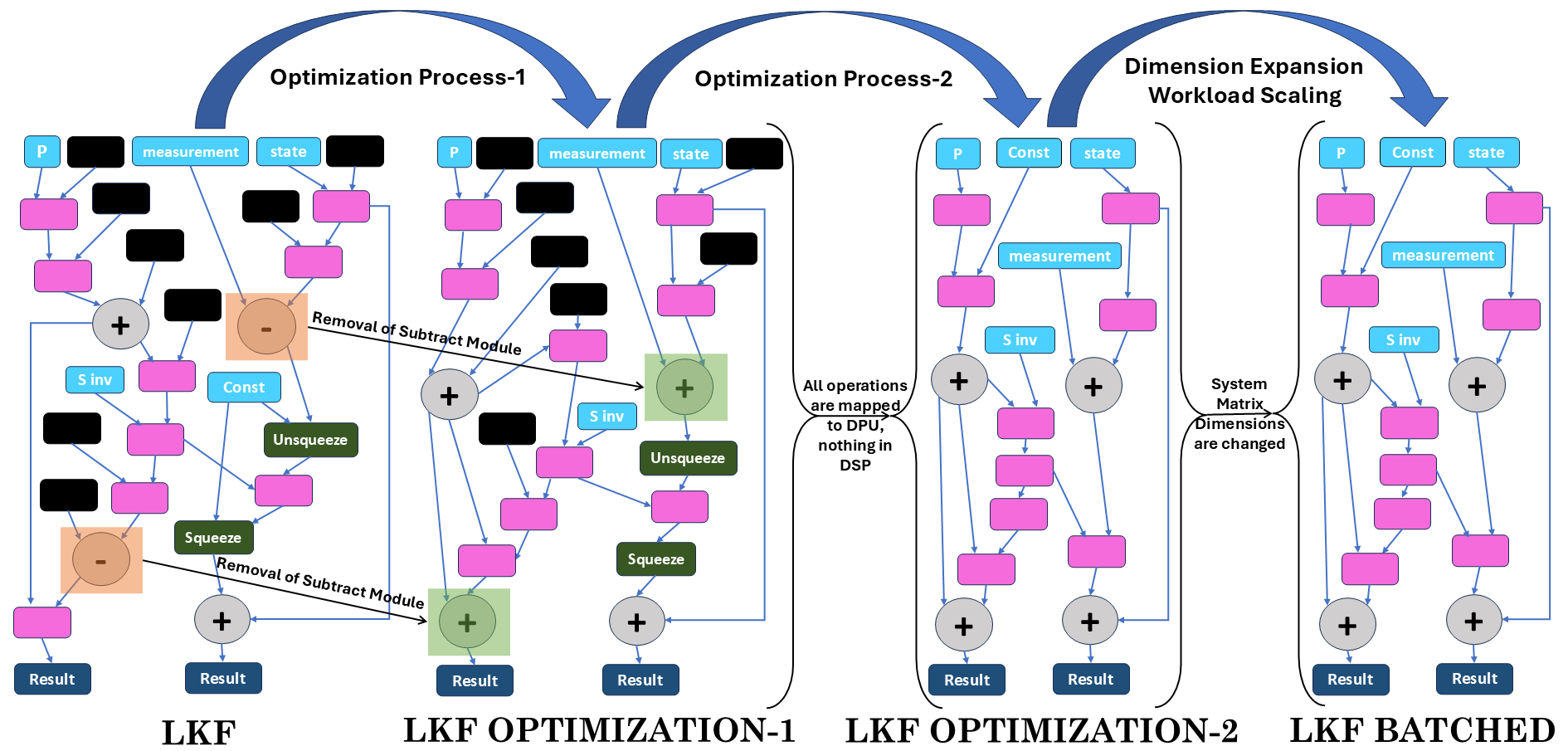}
\end{minipage}
\vspace{0.05cm}
\begin{minipage}{\columnwidth}
    \centering
    \includegraphics[width=\columnwidth]{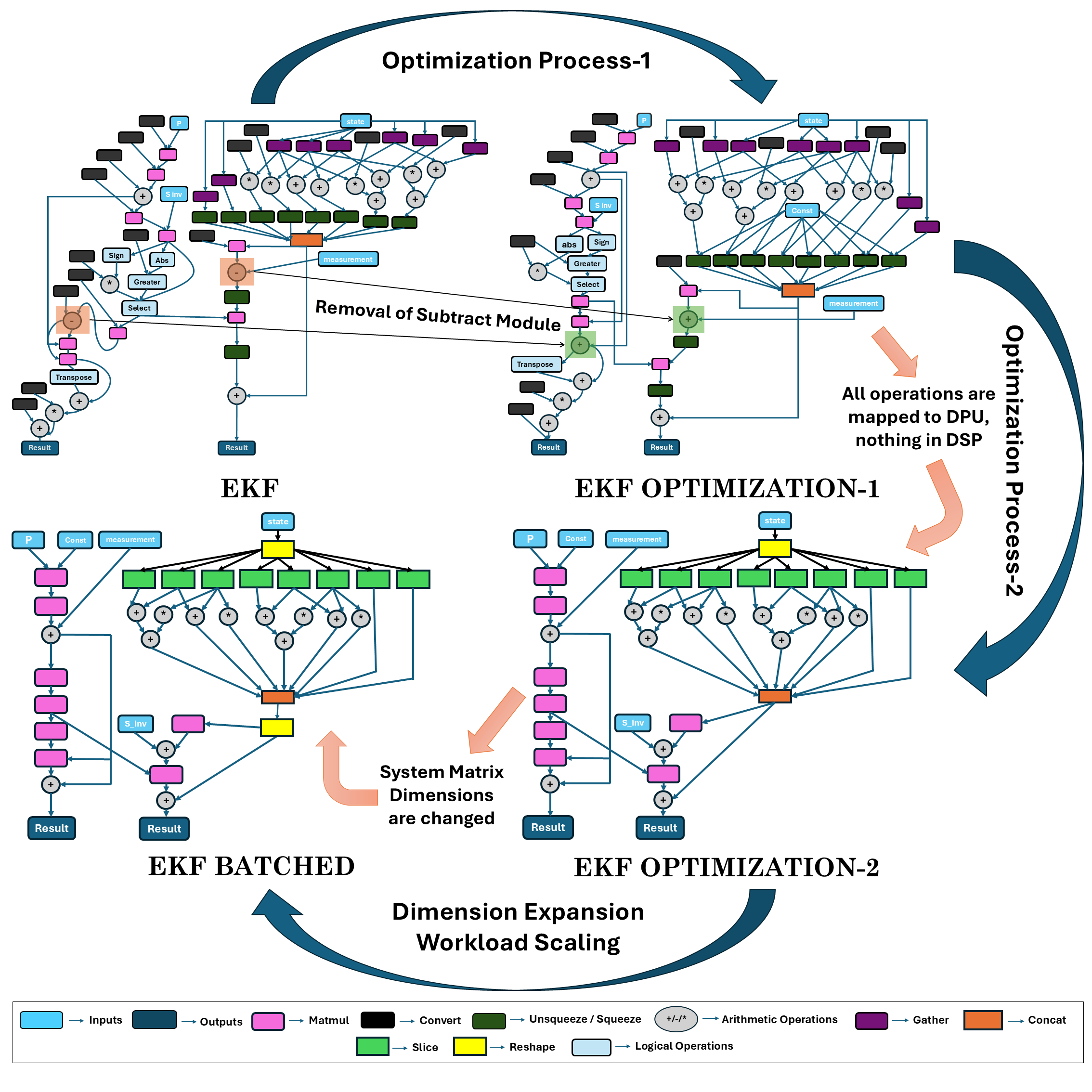}
\end{minipage}
\caption{Netron graphs of the LKF (top) and EKF (bottom) inference pipelines across four optimization stages. \textit{Baseline} retains DSP-bound Subtract and dynamic-shape nodes (highlighted); \textit{Optimization-1} replaces the Subtract module with an Add over a precomputed $\mathbf{H}_{\text{neg}}$; \textit{Optimization-2} fuses static tensor shapes and removes runtime Transposes, leaving the graph entirely on the DPU; \textit{Batched} expands the system-matrix dimensions to process $N=200$ filters per inference call.}
\label{fig:netron_single_column}
\end{figure}

In the baseline graphs of Fig.~\ref{fig:netron_single_column}, the OpenVINO compiler falls back to the DSP for the subtraction in the innovation $\mathbf{y}_k = \mathbf{z}_k - \mathbf{H}\hat{\mathbf{x}}_{k|k-1}$ and to DMA helpers for the dynamic Reshape, Unsqueeze, and Gather nodes introduced by the default batch axis. We measure that these scalar and memory-management nodes account for roughly 15\% of execution time on the LKF baseline (Subtract) and a further 12\% on the EKF baseline (FP32-to-FP16 Convert and other control ops). Concretely, Optimization-1 removes the explicit Subtract module by precomputing a negative-projection matrix; Optimization-2 collapses the remaining dynamic shapes and Transposes so that no DSP-bound node survives; and the Batched stage expands the system matrices into block-diagonal form so that a single inference processes $N=200$ independent filters in parallel. We describe each rewrite in the next three subsections.

\subsection{Algebraic Reformulation: Subtract Elimination}
We address the three sources of DPU fall-back identified in Section~IV-A in turn, starting with the dominant Subtract op. The innovation update can be reformulated by absorbing the sign of the observation projection into a constant tensor. Defining a precomputed matrix $\mathbf{H}_{\text{neg}} = -\mathbf{H}$ at model initialization transforms the innovation from $\mathbf{y}_k = \mathbf{z}_k - \mathbf{H}\hat{\mathbf{x}}_{k|k-1}$ into $\mathbf{y}_k = \mathbf{z}_k + \mathbf{H}_{\text{neg}}\hat{\mathbf{x}}_{k|k-1}$, which is a single GEMM followed by an element-wise Add, both DPU-native. Crucially, $\mathbf{H}_{\text{neg}}$ is folded into the ONNX graph as a constant and carries no runtime cost. The same rewrite applies to every other subtraction in the recursion (state-error and covariance update), so the entire critical path of the LKF and EKF reduces to GEMMs and Adds, as visible in the Optimization-1 column of Fig.~\ref{fig:netron_single_column}.

\subsection{Static Tensor Fusion for Pure-DPU Execution}
Two further rewrites remove the residual DSP and DMA operations visible in the third column of Fig.~\ref{fig:netron_single_column}. First, we lower the dynamic batch axis $[B, \cdot]$ that the ONNX exporter inserts by default to static 1-D and 2-D shapes $[\text{dim}]$; this eliminates the runtime Unsqueeze, Squeeze, and Reshape nodes that OpenVINO had been dispatching to the DSP for shape bookkeeping. Second, we precompute every transposed matrix that appears in the recursion ($\mathbf{F}^T$, $\mathbf{H}^T$, and $\mathbf{H}_{\text{neg}}^T$) and fold them into the graph as constants, so no runtime Transpose op survives. After this stage, the LKF and EKF inference graphs consist exclusively of DPU-native MatMul, Add, and a small number of Concat primitives, as confirmed by the Perfetto traces of Fig.~\ref{fig:perfetto}.

\subsection{Block-Diagonal Batched Parallelization}
While Sections~IV-B and IV-C eliminate DSP fall-back for a single filter, MOT workloads require many filters to run concurrently. For $N$ independent filters, we therefore expand each per-filter matrix into a block-diagonal $(Nn)\times(Nn)$ system matrix, where $n$ is the per-filter state dimension ($n=6$ for our LKF and $n=8$ for our EKF). The measurement, noise, and gain matrices are expanded identically. As a result, this restructuring packs $N$ uncoupled filters into a single NPU inference call: the DPU's MAC pipeline is saturated because the GEMM operands are now wide enough to amortize per-call dispatch overhead, while the block-diagonal sparsity guarantees that filters do not cross-couple numerically. All batched results in Table~\ref{tab:performance} use $N=200$.

\section{Experimental Setup}
\label{sec:exp}
We measure each filter variant of Section~\ref{sec:method} on representative AI-PC hardware, profiling latency, power, and dynamic energy across all three compute targets in the SoC.

\textbf{Hardware platforms.} We evaluate KATANA on two AI-PC class platforms: the Intel Core Ultra Series 1 and Series 2 reference systems. The CPU, GPU, and NPU are measured on the same chassis under the default thermal policy, so that platform variation is the only difference across compute targets.

\textbf{Toolchain.} All filters are authored in PyTorch, exported to ONNX, and compiled to each target backend through Intel OpenVINO 2024.5 \cite{b8}. We use FP16 precision uniformly across CPU, GPU, and NPU to isolate architectural effects from numerical-format effects.

\textbf{Profiling.} Per-operator latency and the DPU/DSP/DMA breakdown shown in Fig.~\ref{fig:perfetto} are captured with the OpenVINO Perfetto trace export. Whole-SoC active power is sampled with the Intel Power Gadget at 100~Hz over the measurement window. Each latency value in Table~\ref{tab:performance} is the mean of 1000 inference iterations after a 100-iteration warm-up.

\textbf{Workloads.} The LKF uses a state of dimension $n=6$ (3-D position and velocity); the EKF uses $n=8$ (constant-turn-rate with acceleration). Single-instance configurations process one filter per inference call; batched configurations process $N=200$ filters in parallel through the block-diagonal expansion of Section~IV-D.

\textbf{End-to-end tracking pipeline.} To validate KATANA on a realistic workload, we deploy it inside a live tracking pipeline derived from the OpenVINO tracking notebooks \cite{b4}. A Haar-cascade detector runs on the CPU and supplies bounding-box centroids as measurements; the NPU-resident LKF and EKF maintain independent state estimates in parallel and feed predicted positions back to the renderer overlay on the next frame. The resulting tracks over a 23-s sequence are shown in Fig.~\ref{fig:tracking_frames}.

\begin{table*}[!t]
\caption{Latency, throughput, power, and dynamic energy of LKF and EKF variants across CPU (C), GPU (G), and NPU (N) on Intel Core Ultra Series 1 and Series 2. Bold marks the best NPU-side cell per (workload, metric).}
\label{tab:performance}
\centering
\setlength{\tabcolsep}{2pt}
\resizebox{\textwidth}{!}{%
\begin{tabular}{@{}l|cccccccccccc|cccccccccccc@{}}
\toprule
& \multicolumn{12}{c|}{\textbf{Series 1}} & \multicolumn{12}{c}{\textbf{Series 2}} \\
\cmidrule(lr){2-13} \cmidrule(lr){14-25}
& \multicolumn{3}{c}{Lat (ms)} & \multicolumn{3}{c}{Thr (FPS)} & \multicolumn{3}{c}{Pwr (W)} & \multicolumn{3}{c|}{Eng (mJ)}
& \multicolumn{3}{c}{Lat (ms)} & \multicolumn{3}{c}{Thr (FPS)} & \multicolumn{3}{c}{Pwr (W)} & \multicolumn{3}{c}{Eng (mJ)} \\
\cmidrule(lr){2-4} \cmidrule(lr){5-7} \cmidrule(lr){8-10} \cmidrule(lr){11-13}
\cmidrule(lr){14-16} \cmidrule(lr){17-19} \cmidrule(lr){20-22} \cmidrule(lr){23-25}
\textbf{Config}
 & C & G & N & C & G & N & C & G & N & C & G & N
 & C & G & N & C & G & N & C & G & N & C & G & N \\
\midrule
LKF              & 0.05 & 0.17 & 0.30 & 15834.40 & 5238.68 & 3096.10 & 28.00 & 28.00 & \textbf{25.92} & 1.40 & 4.76 & 7.77 & 0.02 & 0.30 & 0.25 & 71753.38 & 96448.02 & 6268.29 & 19.98 & 17.11 & \textbf{10.42} & 0.40 & 5.13 & 2.61 \\
LKF OPT 1        & 0.04 & 0.21 & 0.30 & 16648.75 & 4299.45 & 3110.31 & 28.00 & 28.00 & \textbf{26.66} & 1.12 & 5.88 & 8.00 & 0.02 & 0.31 & 0.19 & 75432.49 & 96555.99 & 8105.27 & 20.01 & 16.99 & \textbf{10.93} & 0.40 & 5.27 & 2.08 \\
LKF OPT 2        & 0.05 & 0.17 & 0.29 & 14367.23 & 5345.74 & 3229.84 & 28.00 & 28.00 & \textbf{27.68} & 1.40 & 4.76 & 8.03 & 0.02 & 0.29 & 0.19 & 76877.72 & 5467.01 & 8099.65 & 19.93 & 11.88 & \textbf{9.71} & 0.40 & 3.45 & 1.84 \\
LKF BATCHED      & 79.40 & 5.15 & 5.86 & 12.38 & 154.19 & 145.18 & 28.00 & 28.01 & \textbf{20.06} & 2223.20 & 144.23 & \textbf{117.55} & 69.05 & 1.77 & 2.96 & 23.75 & 640.10 & 408.73 & 17.41 & 25.61 & \textbf{14.05} & 1202.16 & 45.33 & \textbf{41.59} \\
\midrule
EKF              & 0.06 & 0.43 & 0.48 & 12819.82 & 2189.93 & 1964.70 & 28.00 & 28.00 & \textbf{23.00} & 1.68 & 12.04 & 11.04 & 0.03 & 0.57 & 0.41 & 66664.12 & 84068.90 & 3913.76 & 20.91 & 18.89 & \textbf{11.31} & 0.63 & 10.77 & 4.64 \\
EKF OPT 1        & 0.06 & 0.45 & 0.47 & 12740.57 & 2087.28 & 1987.45 & 28.00 & 28.00 & \textbf{24.84} & 1.68 & 12.60 & 11.67 & 0.04 & 1.04 & 0.34 & 22850.51 & 94521.34 & 4650.89 & 22.00 & 15.75 & \textbf{7.47} & 0.88 & 16.38 & 2.54 \\
EKF OPT 2        & 0.05 & 0.25 & 0.29 & 13999.15 & 3670.79 & 3278.13 & 28.00 & 28.00 & \textbf{25.28} & 1.40 & 7.00 & 7.33 & 0.03 & 0.38 & 0.18 & 58684.68 & 3078.13 & 7994.73 & 22.10 & 19.15 & \textbf{12.77} & 0.66 & 7.28 & 2.30 \\
EKF BATCHED      & 122.27 & 15.82 & \textbf{10.84} & 8.06 & 55.09 & \textbf{78.51} & 28.00 & 28.00 & \textbf{19.54} & 3423.56 & 442.96 & \textbf{211.81} & 146.28 & 2.68 & 4.98 & 10.93 & 403.91 & 223.35 & 22.26 & 23.43 & \textbf{13.43} & 3256.46 & 62.79 & \textbf{66.88} \\
\bottomrule
\end{tabular}%
}
\end{table*}

\vspace{-5pt}
\section{Results and Discussion}
\label{sec:results}
We characterize KATANA along four axes: the per-stage compute breakdown that confirms DPU saturation (Fig.~\ref{fig:perfetto}); the latency and throughput scaling with workload density (Table~\ref{tab:performance}); the sustained power and dynamic energy budgets; and the end-to-end tracking demonstration on live video (Fig.~\ref{fig:tracking_frames}). Each axis verifies one design hypothesis from Section~\ref{sec:method} in turn.

\subsection{Compute-Stage Breakdown}
\begin{figure}[t]
\centering
\begin{minipage}{\columnwidth}
  \centering
  \includegraphics[width=\columnwidth]{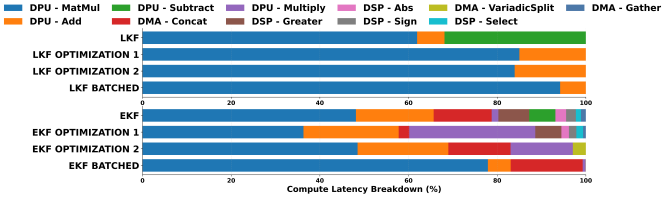}
\end{minipage}
\vspace{0.05cm}
\vspace{0.05cm}
\caption{Per-operator compute-latency breakdown reported by the OpenVINO Perfetto trace for (a) LKF and (b) EKF across the four optimization stages of Section~\ref{sec:method}. Each successive stage migrates work off DSP and DMA paths and onto the DPU; the optimized batched configurations are dominated by DPU-MatMul, confirming the saturation of the matrix engine.}
\label{fig:perfetto}
\end{figure}

Fig.~\ref{fig:perfetto} reports the per-operator NPU compute breakdown across the four stages of Section~\ref{sec:method}. For the LKF, successive stages eliminate the baseline DPU Subtract bar and the residual non-MAC tail, leaving the Batched configuration with $\approx$95\% of execution in DPU MatMul. The EKF baseline scatters work across several DSP and DMA tail ops that together account for roughly half of inference time; after the three rewrites, the batched EKF reaches $\approx$80\% DPU MatMul with the rest in DPU Add and DMA Concat. The Netron-level transformations of Fig.~\ref{fig:netron_single_column} thus translate directly into measurable DPU occupancy.

\subsection{Latency and Throughput Scaling}
Table~\ref{tab:performance} lists latency, throughput, power, and dynamic energy for every (filter, platform, configuration) combination; two regimes emerge.

\textit{Single-instance regime.} At one filter per inference call, the CPU wins on raw latency (down to 0.02~ms on the Series 2) because the workload is too small to amortize NPU dispatch overhead; the NPU still meets any real-time budget at 0.18--0.30~ms but is not yet exercising its advantage.

\textit{Batched regime ($N{=}200$).} Once the workload is dense enough to keep the matrix engine busy, the NPU is the throughput-best target on the Series~1 SoC for the EKF (\textbf{78.5~FPS} vs.\ 8.1 CPU / 55.1 GPU). The Series~2 nearly triples NPU throughput to \textbf{223.35~FPS} (EKF) and pushes the LKF to \textbf{408.73~FPS}, validating the block-diagonal batching design of Section~IV-D. The Series~2 GPU reaches higher peak FPS through race-to-idle execution but at substantially higher sustained power, as quantified next.

\subsection{Power and Energy Efficiency}
\label{subsec:power}
For the EKF $N{=}200$ workload, the Series~2 NPU sustains only \textbf{13.43~W} against 22--25~W for the CPU/GPU on the same chassis, and the corresponding \textbf{66.88~mJ} of dynamic energy is a \textbf{97.9\%} reduction relative to the Series~2 CPU on the same workload (Table~\ref{tab:performance}). The GPU reaches competitive throughput through race-to-idle execution but at a peak power that fanless, battery-bound edge platforms cannot sustain. The NPU's advantage is therefore sustained low power, not peak FLOPs: precisely the regime in which defense and mobile tracking platforms operate.

\begin{figure}[t]
\vspace{-0.15cm}
\centering
\setlength{\tabcolsep}{1pt}
\begin{tabular}{cccccc}
\includegraphics[width=0.155\columnwidth]{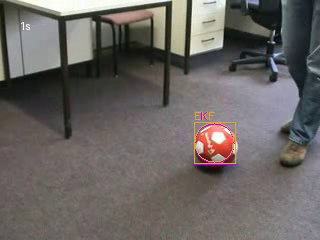} &
\includegraphics[width=0.155\columnwidth]{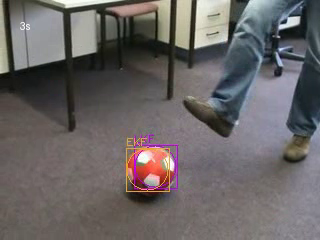} &
\includegraphics[width=0.155\columnwidth]{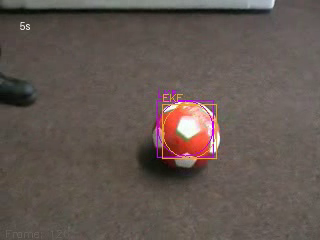} &
\includegraphics[width=0.155\columnwidth]{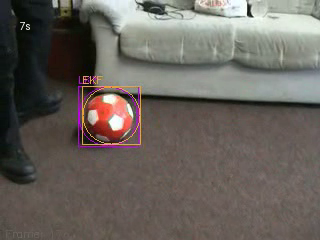} &
\includegraphics[width=0.155\columnwidth]{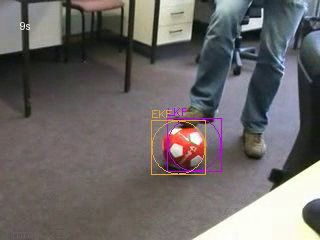} &
\includegraphics[width=0.155\columnwidth]{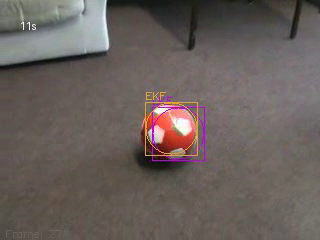} \\[-0.5mm]
\includegraphics[width=0.155\columnwidth]{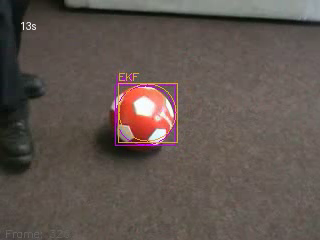} &
\includegraphics[width=0.155\columnwidth]{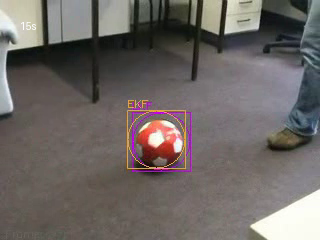} &
\includegraphics[width=0.155\columnwidth]{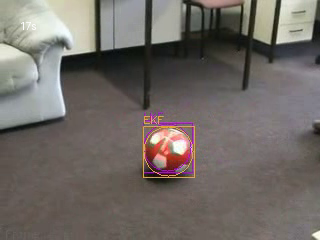} &
\includegraphics[width=0.155\columnwidth]{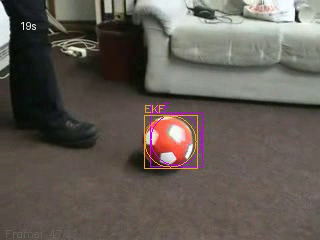} &
\includegraphics[width=0.155\columnwidth]{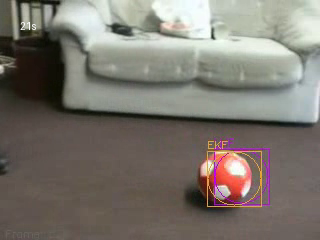} &
\includegraphics[width=0.155\columnwidth]{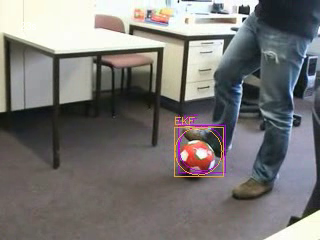}
\end{tabular}
\vspace{0.05cm}
\caption{Frames extracted at odd timestamps from the live KATANA tracking sequence with the LKF and EKF running on the Intel Core Ultra Series 2 NPU. Both filters maintain a stable track through scale and motion changes across the 23-s sequence.}
\label{fig:tracking_frames}
\vspace{0.05cm}
\end{figure}

\subsection{Real-Time Tracking on Live Video}
Fig.~\ref{fig:tracking_frames} shows snapshots from the NPU-accelerated tracking pipeline of Section~\ref{sec:exp}, sampled at odd timestamps over a 23-s sequence. Both filters lock onto the target on the first frame and follow it through scale and motion changes without visible drift. On the Series~2, the optimized single-instance LKF and EKF take \textbf{0.19~ms} and \textbf{0.18~ms} per update, occupying \textbf{$<$1\%} of the 33~ms frame budget at 30~FPS and leaving the rest available to the Haar-cascade detector, a future re-identification head, or other concurrent SoC workloads. The end-to-end demonstration confirms that an NPU-resident KF can serve as the always-on tracking engine without monopolizing the platform.

\vspace{-5pt}
\section{Conclusion}
KATANA delivers the first end-to-end mapping and cross-platform (CPU, GPU, NPU) characterization of the LKF and EKF on commercial AI-PC silicon. Three algebraic rewrites (subtract-to-add reformulation, static-shape fusion, and block-diagonal batching) move 100\% of the recursion onto the DPU, so the optimized pipeline sustains $>$200~FPS multi-filter throughput at sub-15~W and cuts dynamic energy by up to \textbf{97.9\%} versus the CPU. Classical signal-processing pipelines can therefore be re-targeted to existing AI-PC NPUs without custom accelerators.
\vspace{-5pt}

\end{document}